\input amstex
\documentstyle{amsppt}
%
\catcode`@=11
\redefine\output@{%
  \def\break{\penalty-\@M}\let\par\endgraf
  \ifodd\pageno\global\hoffset=105pt\else\global\hoffset=8pt\fi  
  \shipout\vbox{%
    \ifplain@
      \let\makeheadline\relax \let\makefootline\relax
    \else
      \iffirstpage@ \global\firstpage@false
        \let\rightheadline\frheadline
        \let\leftheadline\flheadline
      \else
        \ifrunheads@ 
        \else \let\makeheadline\relax
        \fi
      \fi
    \fi
    \makeheadline \pagebody \makefootline}%
  \advancepageno \ifnum\outputpenalty>-\@MM\else\dosupereject\fi
}
\catcode`\@=\active
\nopagenumbers
\def\tr{\operatorname{tr}}
\def\negskp{\hskip -2pt}

\def\blue#1{#1}
\catcode`#=11\def\diez{#}\catcode`#=6
\catcode`_=11\def\podcherkivanie{_}\catcode`_=8
\def\mycite#1{\cite{\blue{#1}}\immediate\special{ps:
     ShrHPSdict begin /ShrBORDERthickness 0 def}}

\def\mytag#1{%
    \tag#1}
\def\mythetag#1{\thetag{\blue{#1}}\immediate\special{ps:
     ShrHPSdict begin /ShrBORDERthickness 0 def}}
\def\myrefno#1{\no#1}
\def\myhref#1#2{\blue{#2}\immediate\special{ps:
     ShrHPSdict begin /ShrBORDERthickness 0 def}}

\def\mytheorem#1{\csname proclaim\endcsname{Theorem #1}}
\def\mythetheorem#1{\blue{#1}\immediate\special{ps:
     ShrHPSdict begin /ShrBORDERthickness 0 def}}
\def\mylemma#1{\csname proclaim\endcsname{Lemma #1}}

\def\mycorollary#1{\csname proclaim\endcsname{Corollary #1}}

\def\mydefinition#1{\definition{Definition #1}}
\def\mythedefinition#1{\blue{#1}\immediate\special{ps:
     ShrHPSdict begin /ShrBORDERthickness 0 def}}

\pagewidth{360pt}
\pageheight{606pt}
\topmatter
\title
Algorithms for laying points optimally on a plane and a circle.
\endtitle
\author
R.~A.~Sharipov
\endauthor
\address 5 Rabochaya street, 450003 Ufa, Russia\newline
\vphantom{a}\kern 12pt Cell Phone: +7-(917)-476-93-48
\endaddress
\email \vtop to 30pt{\hsize=280pt\noindent
\myhref{mailto:r-sharipov\@mail.ru}
{r-sharipov\@mail.ru}\newline
\myhref{mailto:R\podcherkivanie Sharipov\@ic.bashedu.ru}
{R\_\hskip 1pt Sharipov\@ic.bashedu.ru}\vss}
\endemail
\urladdr
\vtop to 20pt{\hsize=280pt\noindent
\myhref{http://www.geocities.com/r-sharipov}
{http:/\negskp/www.geocities.com/r-sharipov}\newline
\myhref{http://www.freetextbooks.boom.ru/index.html}
{http:/\negskp/www.freetextbooks.boom.ru/index.html}\vss}
\endurladdr
\abstract
    Two averaging algorithms are considered which are intended
for choosing an optimal plane and an optimal circle approximating
a group of points in three-dimensional Euclidean space.
\endabstract
\subjclassyear{2000}
\subjclass 62H35, 62P30, 68W25\endsubjclass
\endtopmatter
\TagsOnRight
\document

\rightheadtext{Algorithms for laying points \dots}
\head
1. Introduction.
\endhead
\parshape 3 0pt 360pt 0pt 360pt 180pt 180pt 
    Assume that in the three-dimensional Euclidean space $\Bbb E$ we 
have a group of points visually resembling a circle (see Fig\.~1.1). 
\vadjust{\vskip 5pt\hbox to 0pt{\kern -30pt
\includegraphics{Circ01.eps}\hss}\vskip -5pt}The problem is 
to find the best plane and the best circle approximating this group of 
points. Any plane in $\Bbb E$ is given by the equation
$$
\hskip -2em
(\bold r,\bold n)=D,
\mytag{1.1}
$$
where $\bold n$ is the normal vector of the  plane and $D$ is some
constant. The vector $\bold r$ in \mythetag{1.1} is the radius-vector
of a point on that plane, while $(\bold r,\bold n)$ is the scalar
pro\-duct of the vectors $\bold r$ and $\bold n$.\par
\parshape 8 180pt 180pt 180pt 180pt 180pt 180pt 180pt 180pt 
180pt 180pt 180pt 180pt 180pt 180pt 0pt 360pt
     Once a plane \mythetag{1.1} is fixed and $\bold r$ is the 
radius-vector of some point on it, a circle on this plane is given by 
the equation
$$
\hskip -2em
|\bold r-\bold R|=\rho.
\mytag{1.2}
$$
Here $\rho$ is the radius of the circle \mythetag{1.2} and $\bold R$
is the radius-vector of its center. Having a group of points $\bold r[1],\,\ldots,\,\bold r[N]$ in $\Bbb E$, our goal is to design an 
algorithm for calculating the parameters $\bold n$, $D$, $\bold R$, 
and $\rho$ in \mythetag{1.1} and \mythetag{1.2} thus defining a plane
and a circle being optimal approximations of our points in some definite
sense.\par
\head
2. Defining an optimal plane.
\endhead
    Assume that $\bold n$ is a unit vector, i\.\,e\, $|\bold n|=1$, 
and assume that we have some plane defined by the equation 
\mythetag{1.1}. \pagebreak Then the distance from the point 
$\bold r[i]$ to this plane is given by the following well-known 
formula:
$$
\hskip -2em
d[i]=\frac{|(\bold r[i],\bold n)-D|}{|\bold n|}
=|(\bold r[i],\bold n)-D|.
\mytag{2.1}
$$
If we denote by $d$ the root of mean square of the quantities 
\mythetag{2.1}, then we have
$$
\hskip -2em
d^{\,2}=\frac{1}{N}\sum^N_{i=1}d[i]^2=\frac{1}{N}\sum^N_{i=1}
|(\bold r[i],\bold n)-D|^2.
\mytag{2.2}
$$
\mydefinition{2.1} A plane given by the formula \mythetag{1.1} with
$|\bold n|=1$ is called an {\it optimal root mean square plane\/} if
the quantity \mythetag{2.2} takes its minimal value.
\enddefinition
    It is easy to see that $d^{\,2}$ in \mythetag{2.2} is a function 
of two parameters: $\bold n$ and $D$. It is a quadratic function of
the parameter $D$. Indeed, we have
$$
\hskip -2em
d^{\,2}=D^2-\frac{2}{N}\sum^N_{i=1}(\bold r[i],\bold n)\,D
+\frac{1}{N}\sum^N_{i=1}(\bold r[i],\bold n)^2.
\mytag{2.3}
$$
The quadratic polynomial in the right hand side of \mythetag{2.3} takes
its minimal value if 
$$
\hskip -2em
D=\frac{1}{N}\sum^N_{i=1}(\bold r[i],\bold n).
\mytag{2.4}
$$
Substituting \mythetag{2.4} back into the formula \mythetag{2.3}, we obtain
$$
\hskip -2em
d^{\,2}=\frac{1}{N}\sum^N_{i=1}(\bold r[i],\bold n)^2
-\left(\frac{1}{N}\sum^N_{i=1}(\bold r[i],\bold n)\right)^{\!\lower 4pt
\hbox{$\ssize 2$}}.
\mytag{2.5}
$$
In the next steps we use some mechanical analogies. If we place unit 
masses $m[i]=1$ at the points $\bold r[1],\,\ldots,\,\bold r[N]$, then 
the vector 
$$
\hskip -2em
\bold r_{\text{cm}}=\frac{1}{N}\sum^N_{i=1}\bold r[i]
\mytag{2.6}
$$
is the radius-vector of the center of mass. In terms of this radius
vector the formula \mythetag{2.6} for $D$ is written as follows:
$$
\hskip -2em
D=(\bold r_{\text{cm}},\bold n).
\mytag{2.7}
$$
Now remember that the inertia tensor for a system of point masses
$m[i]=1$ is defined as a quadratic form given by the formula:
$$
I(\bold n,\bold n)=\sum^N_{i=1}|\bold r[i]|^2\,|\bold n|^2-
\sum^N_{i=1}(\bold r[i],\bold n)^2
\mytag{2.8}
$$
(see \mycite{1} for more details). \pagebreak We shall take the 
inertia tensor relative to the center of mass. Therefore, we 
substitute $\bold r[i]-\bold r_{\text{cm}}$ for $\bold r[i]$ into 
the formula \mythetag{2.8}. As a result we get the following 
expression for $I(\bold n,\bold n)$:
$$
\hskip -2em
I(\bold n,\bold n)=\sum^N_{i=1}|\bold r[i]-\bold r_{\text{cm}}|^2
\,|\bold n|^2-\sum^N_{i=1}(\bold r[i]-\bold r_{\text{cm}},\bold n)^2.
\mytag{2.9}
$$
Each quadratic form in a three-dimensional Euclidean space has $3$ 
scalar invariants. One of them is trace the invariant. In the case
of the quadratic form \mythetag{2.9}, the trace invariant is given 
by the following formula:
$$
\hskip -2em
\tr(I)=2\sum^N_{i=1}|\bold r[i]-\bold r_{\text{cm}}|^2.
\mytag{2.10}
$$
Combining \mythetag{2.9} and \mythetag{2.10}, we write
$$
\hskip -2em
I(\bold n,\bold n)=\frac{\tr(I)}{2}\,|\bold n|^2
-\sum^N_{i=1}(\bold r[i]-\bold r_{\text{cm}},\bold n)^2.
\mytag{2.11}
$$
Taking into account the formula \mythetag{2.6}, we transform
\mythetag{2.11} as follows:
$$
\hskip -2em
I(\bold n,\bold n)=\frac{\tr(I)}{2}\,|\bold n|^2
-\sum^N_{i=1}(\bold r[i],\bold n)^2
+N\,(\bold r_{\text{cm}},\bold n)^2.
\mytag{2.12}
$$
Comparing \mythetag{2.12} with \mythetag{2.5} and again taking
into account \mythetag{2.6}, we get
$$
\hskip -2em
d^{\,2}=\frac{\tr(I)}{2\,N}\,|\bold n|^2
-\frac{I(\bold n,\bold n)}{N}.
\mytag{2.13}
$$
The formula \mythetag{2.13} means that $d^{\,2}$ is a quadratic
form similar to the inertia tensor. We call it the {\it non-flatness
form} and denote $Q(\bold n,\bold n)$:
$$
\hskip -2em
\gathered
Q(\bold n,\bold n)=\frac{\tr(I)}{2\,N}\,|\bold n|^2
-\frac{I(\bold n,\bold n)}{N}=\\
=\frac{1}{N}\sum^N_{i=1}(\bold r[i],\bold n)^2
-\left(\frac{1}{N}\sum^N_{i=1}(\bold r[i],\bold n)
\right)^{\!\lower 4pt\hbox{$\ssize 2$}}.
\endgathered
\mytag{2.14}
$$
Like the inertia form \mythetag{2.9}, the non-flatness form 
\mythetag{2.14} is positive, i\.\,e\.
$$
Q(\bold n,\bold n)\geqslant 0\text{\ \  for \ }\bold n\neq 0.
$$
If the inertia tensor is brought to its primary axes, i\.\,e\.
if it is diagonalized in some orthonormal basis, then the form
\mythetag{2.14} diagonalizes in the same basis.
\mytheorem{2.1} A plane is an optimal root mean square plane
for a group of points if and only if it passes through the
center of mass of these points and if its normal vector $\bold n$
is directed along a primary axis of the non-flatness form $Q$ of
these points corresponding to its minimal \pagebreak eigenvalue.
\endproclaim
    The proof is derived immediately from the 
definition~\mythedefinition{2.1} due to the formula \mythetag{2.7} 
and the formula $d^{\,2}=Q(\bold n,\bold n)$.
\mytheorem{2.2} An optimal root mean square plane for a group of 
points is unique if and only if the minimal eigenvalue 
$\lambda_{\text{min}}$ of their non-flatness form $Q$ is distinct
from two other eigenvalues, i\.\,e\.
$\lambda_{\text{min}}=\lambda_1<\lambda_2$ and
$\lambda_{\text{min}}=\lambda_1<\lambda_3$.
\endproclaim
\head
3. Defining an optimal circle.
\endhead
    Having found an optimal root mean square plane for the 
points $\bold r[1],\,\ldots,\,\bold r[N]$, we can replace them by
their projections onto this plane:
$$
\hskip -2em
\bold r[i]\ \mapsto\ \bold r[i]-((\bold r[i],\bold n)-D)\ \bold n.
\mytag{3.1}
$$
Our next goal is to find an optimal circle approximating a group of
points lying on some plane \mythetag{1.1}. Let $\bold r[1],\,\ldots,
\,\bold r[N]$ be their radius-vectors. The deflection of the point
$\bold r[i]$ from the circle \mythetag{1.2} is characterized by the
following quantity:
$$
\hskip -2em
d[i]=||\bold r[i]-\bold R|^2-\rho^2|.
\mytag{3.2}
$$
Like in the case of \mythetag{2.1}, we denote by $d$ the root mean
square of the quantities \mythetag{3.2}. Then we get the following
formula:
$$
\hskip -2em
d^{\,2}=\frac{1}{N}\sum^N_{i=1}d[i]^2=\frac{1}{N}\sum^N_{i=1}
(|\bold r[i]-\bold R|^2-\rho^2)^2.
\mytag{3.3}
$$
The quantity $d^{\,2}$ in \mythetag{3.3} is a function of two parameters:
$\bold R$ and $\rho^2$. With respect to $\rho^2$ it is a quadratic
polynomial. Indeed, we have
$$
\hskip -2em
d^{\,2}=(\rho^2)^2
-\frac{2\,\rho^2}{N}\sum^N_{i=1}|\bold r[i]-\bold R|^2
+\frac{1}{N}\sum^N_{i=1}|\bold r[i]-\bold R|^4.
\mytag{3.4}
$$
Being a quadratic polynomial of $\rho^2$, the quantity $d^{\,2}$ takes
its minimal value for
$$
\hskip -2em
\rho^2=\frac{1}{N}\sum^N_{i=1}|\bold r[i]-\bold R|^2.
\mytag{3.5}
$$
Substituting \mythetag{3.5} back into the formula \mythetag{3.4}, we 
derive
$$
\hskip -2em
d^{\,2}=\frac{1}{N}\sum^N_{i=1}|\bold r[i]-\bold R|^4
-\left(\frac{1}{N}\sum^N_{i=1}|\bold r[i]-\bold R|^2\right)^{\!\lower 
4pt\hbox{$\ssize 2$}}.
\mytag{3.6}
$$
Upon expanding the expression in the right hand side of the formula
\mythetag{3.6} we need to perform some simple, but rather huge 
calculations. As result we get
$$
\allowdisplaybreaks
\gather
d^{\,2}=\frac{1}{N}\sum^N_{i=1}|\bold r[i]|^4
-\left(\frac{1}{N}\sum^N_{i=1}|\bold r[i]|^2\right)^{\!\lower 
4pt\hbox{$\ssize 2$}}-\frac{4}{N}\sum^N_{i=1}|\bold r[i]|^2
\,(\bold r[i],\bold R)\,+\\
+\ 4\left(\frac{1}{N}\sum^N_{i=1}|\bold r[i]|^2\right)
\!\left(\frac{1}{N}\sum^N_{i=1}(\bold r[i],\bold R)\right)
+\frac{4}{N}\sum^N_{i=1}(\bold r[i],\bold R)^2
-4\left(\frac{1}{N}\sum^N_{i=1}(\bold r[i],\bold R)\right)^{\!\lower 
4pt\hbox{$\ssize 2$}}.
\endgather
$$
We see that the above expression is not higher than quadratic with 
respect to $\bold R$. The fourth order terms and the cubic terms are 
canceled. Note also that the quadratic part of the above expression
is determined by the form $Q$ considered in previous section. For
this reason we write $d^{\,2}$ as 
$$
\hskip -2em
d^{\,2}=4\,Q(\bold R,\bold R)-4\,(\bold L,\bold R)+M.
\mytag{3.7}
$$
The vector $\bold L$ and the scalar $M$ in \mythetag{3.7} are given 
by the following formulas:
$$
\gather
\hskip -2em
\bold L=\frac{1}{N}\sum^N_{i=1}|\bold r[i]|^2\,(\bold r[i]
-\bold r_{\text{cm}}),
\mytag{3.8}\\
M=\frac{1}{N}\sum^N_{i=1}|\bold r[i]|^4
-\left(\frac{1}{N}\sum^N_{i=1}|\bold r[i]|^2\right)^{\!\lower 
4pt\hbox{$\ssize 2$}}.
\mytag{3.9}
\endgather
$$
The quantity $d^{\,2}$ takes its minimal value if and only if
$\bold R$ satisfies the equation
$$
\hskip -2em
2\bold Q(\bold R)=\bold L,
\mytag{3.10}
$$
where $\bold Q$ is the symmetric linear operator associated with
the form $Q$ through the standard Euclidean scalar product. The
equality 
$$
(\bold Q(\bold X),\bold Y)=Q(\bold X,\bold Y),
$$
which should be fulfilled for arbitrary two vectors $\bold X$ and 
$\bold Y$, is a formal definition of the operator $\bold Q$
(see \mycite{2} for more details).\par
    In general case the operator $\bold Q$ is non-degenerate. Hence,
$\bold R$ does exist and uniquely fixed by the equation \mythetag{3.10}.
However, if the points $\bold r[1],\,\ldots,\,\bold r[N]$ are laid
onto the plane \mythetag{1.1} by means of the projection procedure
\mythetag{3.1}, then the operator $\bold Q$ is degenerate. Moreover,
one can prove the following theorem.
\mytheorem{3.1}The non-flatness form $Q$ and its associated operator 
$\bold Q$ are degenerate if and only if the points $\bold r[1],\,
\ldots,\,\bold r[N]$ lie on some plane.
\endproclaim
     In this flat case provided by the theorem~\mythetheorem{3.1} 
one should move the origin to that plane where the points $\bold r[1],
\,\ldots,\,\bold r[N]$ lie and treat their radius-vectors 
as two-dimensional vectors. Then, using \mythetag{2.14}, \mythetag{3.8}, 
and \mythetag{3.9}, one should rebuild the two-dimensional versions
of the non-flatness form $Q$, its associated operator $\bold Q$ and
the parameters $\bold L$ and $M$. If again the two-dimensional
non-flatness form is degenerate, this case is described by the following
theorem.
\mytheorem{3.2} The two-dimensional non-flatness form $Q$ and its 
associated operator $\bold Q$ are degenerate if and only if all of
the points $\bold r[1],\,\ldots,\,\bold r[N]$ lie on some straight
line.
\endproclaim
In this very special case we say that straight line approximation for
the points $\bold r[1],\,\ldots,\,\bold r[N]$ is more preferable than 
the circular approximation. Note that the same decision can be made
in some cases even if the points $\bold r[1],\,\ldots,\,\bold r[N]$
do not lie on one straight line exactly. If two eigenvalues of the
three-dimensional non-flatness form $Q$ are sufficiently small, 
i\.\,e\. if they both are much smaller than the third eigenvalue
of this form, then we can say that 
$$
\xalignat 2
&\lambda_{\text{min}}\approx\lambda_1,
&&\lambda_{\text{min}}\approx\lambda_2.
\endxalignat
$$
Taking two eigenvectors $\bold n_1$ and $\bold n_2$ of the form $Q$ corresponding to the eigenvalues $\lambda_1$ and $\lambda_2$, we 
define two planes
$$
\xalignat 2
&\hskip -2em
(\bold r,\bold n_1)=D_1,
&&(\bold r,\bold n_2)=D_2.
\mytag{3.11}
\endxalignat
$$
The constants $D_1$ and $D_2$ in \mythetag{3.11} are given by the
formula \mythetag{2.7}. The intersection of two planes \mythetag{3.11}
yields a straight line being the optimal straight line approximation 
for the points $\bold r[1],\,\ldots,\,\bold r[N]$ in this case.
\head
4. Acknowledgments.
\endhead
    The idea of this paper was induced by some technological problems
suggested to me by O.~V.~Ageev. I am grateful to him for that.

\enddocument
\end